\documentclass[12pt]{article}
\pdfoutput=1
\usepackage{draft}
\usepackage{comment}
\usepackage[normalem]{ulem}
\usepackage{hyperref}
\hypersetup{
 colorlinks=true,
 urlcolor=blue,
 anchorcolor=blue,
 citecolor=blue,
 filecolor=blue,
 linkcolor=blue,
 menucolor=blue,
 pagecolor=blue,
 linktocpage=true,
}
\usepackage{graphicx,color,subfig}
\usepackage{cite}
\usepackage{empheq} 
\usepackage[font=small,labelfont=bf]{caption} 

\usepackage[usenames,dvipsnames,table]{xcolor}
\graphicspath{{./figures/}}
\usepackage{tikz}
\usepackage{empheq}

\def\cc{\cite}

\def\foot{\footnote}

\usepackage{scalerel}
\usepackage{stackengine,wasysym}

\DeclareFontFamily{OT1}{pzc}{}
\DeclareFontShape{OT1}{pzc}{m}{it}{<-> s * [1.10] pzcmi7t}{}
\DeclareMathAlphabet{\mathpzc}{OT1}{pzc}{m}{it}

\definecolor{vert}{rgb}{0.1367 0.543 0.1367}

\def\({\left(}
\def\){\right)}

\usepackage{chngcntr}
\counterwithout{equation}{section}

\begin{document}

\unitlength = .8mm

\begin{titlepage}

	  \begin{flushright}
	 \end{flushright} 

\begin{center}

 \hfill \\
 \hfill \\

\title{Snowmass White Paper: Bootstrapping String Theory}

~\vskip 0.01 in

\author{
	Rajesh Gopakumar,$^{1}$ Eric Perlmutter,$^{2}$ Silviu S.~Pufu,$^{3}$ and Xi Yin$^{4}$ }

${}^1$\emph{\small International Centre for Theoretical Sciences-TIFR, \\
Shivakote, Hesaraghatta Hobli, Bengaluru North, 560 089, India}\\
\vskip .1 in
${}^2$\emph{\small Universit\'e Paris-Saclay, CNRS, CEA, Institut de Physique Th\'eorique, \\91191, Gif-sur-Yvette, France
\\
\vskip .1 in
${}^3$Joseph Henry Laboratories, Princeton University, Princeton, NJ 08544, USA
\\
\vskip .1 in
${}^4$Jefferson Physical Laboratory, Harvard University, 
Cambridge, MA 02138 USA}
\\

\email{rajesh.gopakumar@icts.res.in, 
	perl@ipht.fr, spufu@princeton.edu,
	xiyin@fas.harvard.edu
}

\end{center}

\vskip .4 in

\abstract{\vskip .1 in \centerline{We discuss progress and prospects in the application of bootstrap methods to string theory.}}

\vfill

\end{titlepage}

\eject

\begingroup

\baselineskip .168 in
\tableofcontents

\endgroup

\section{Introduction}

The Veneziano amplitude, that birthed string theory, itself originated from the analytic S-matrix bootstrap philosophy of imposing strong analyticity constraints on scattering amplitudes to capture general principles of locality, unitarity, and causality. A dynamical explanation of this remarkable formula and its generalizations followed subsequently from a two-dimensional field theory describing the worldsheet of a relativistic string. Later, in the early eighties, it was the success of the conformal bootstrap applied to two-dimensional CFTs which spurred the exciting progress, rooted in the worldsheet, that expanded our understanding of the space of string theory vacua. Thus, in some ways, string theory and bootstrap ideas, both in spacetime and on the worldsheet, are joined at the hip. 

Fast-forwarding, we once again find ourselves in the midst of a period when bootstrap ideas have not only had a lot of fresh success with general CFTs or QFTs, but have also begun to fruitfully and nontrivially feed into string theory. The goal of this white paper is to articulate some important elements of this connection, reviewing progress of the last decade with an eye toward what we view as the most important avenues for future research.\footnote{We hasten to state that the narrow remit of this paper does not convey the full scope of progress in the conformal bootstrap program in recent years, nor do we claim to summarize the status and many interesting open questions of string theory {\it per se}.}

In broad terms, this re-engagement of string theory with the bootstrap has two strands once again. The dominant one has been a spacetime one, primarily through the AdS/CFT correspondence. Success with bootstrapping certain CFT$_d$ correlation functions in appropriate large $N$ regimes holographically constrains scattering amplitudes in AdS$_{d+1}$ which, by taking a suitable limit, go over to string scattering amplitudes in flat space. Here both numerical and analytic bootstrap techniques (often in tandem with other tools like supersymmetric localization) play a crucial role, as we summarize in Section~\ref{AMPLITUDES}. One of the highlights here is the determination of  several leading terms in the low-energy (derivative) expansion of string theory/M-theory through this holographic bootstrap. The other natural domain of the conformal bootstrap is to constrain the space of CFTs. In the holographic context, this provides a rigorous approach to charting the space of theories of quantum gravity and, in turn, the existential questions that are difficult to settle with direct computations. An example is the question of existence of scale separated AdS vacua (i.e.~with the non-AdS directions being small), which is important for understanding whether low energy constructions of such vacua do indeed lift to the full string theory. The latter are also often the crucial intermediate stages in constructions of metastable de Sitter vacua. Questions such as these translate, via the AdS/CFT dictionary, into whether there are CFTs with sufficiently sparse primary spectra at low spacetime spin. We discuss these and related topics in Section~\ref{LANDSCAPE}. 

The second strand, which thus far has not made a prominent appearance, is to bring bootstrap ideas to bear on the worldsheet description of the much more general class of string vacua that the AdS/CFT correspondence brings to the forefront. In Section \ref{WORLDSHEET}, we discuss the subtleties that such a worldsheet description will necessarily face, which would probably require us to widen the class of theories beyond what we are conventionally familiar with. Having such a description is important for concrete problems like a first principles description of confining string backgrounds. More broadly, we seek techniques to perform worldsheet calculations of the AdS ampltudes that are indirectly obtained through the holographically dual CFT\@. We outline some ideas on how one might derive mileage from ``dually bootstrapping" the spacetime CFT and worldsheet descriptions.  

We conclude this paper in Section \ref{POST} with a more speculative list of stringy contexts, accompanied by an artistic rendition of the state of the field, where the bootstrap philosophy can potentially be usefully applied. As a counterpoint to that, perhaps, we end this introduction with a list of some very concrete goals, which we contextualize in the main text, that we feel can be tackled in the short-to-medium term using bootstrap methods, as a natural growth of the approaches summarized here:

\begin{itemize}
\item
Obtain the $D^8R^4$ terms in the type II string theory and M-theory effective actions.
\item
Write down the Virasoro-Shapiro amplitude for $AdS_5 \times S^5$.
\item
Definitively rule in/out scale-separated AdS vacua using large $N$ conformal bootstrap.
\item
Bootstrap the worldsheet description of weakly-coupled $\mathcal{N}=4$ super-Yang-Mills theory.  
\end{itemize}

\section{Bootstrapping string theory amplitudes}
\label{AMPLITUDES}

Some of the most fundamental observables in theories of quantum gravity in asymptotically flat spacetime are the scattering amplitudes of gravitons. In string theory, the worldsheet formalism produces, in principle, the complete perturbative expansion of closed string amplitudes \cite{Friedan:1985ge, Green:1987sp, Green:1987mn, DHoker:1988pdl, Polchinski:1998rq, Polchinski:1998rr, Berkovits:2000wm, Berkovits:2006vi, Sen:2014pia, Sen:2015hia, deLacroix:2017lif}, in particular of the gravitons and their superpartners,  and it also captures certain non-perturbative corrections \cite{Polchinski:1994fq}. In practice, however, explicit results of string amplitudes are only available up to two-loop order in perturbation theory \cite{DHoker:2001kkt, DHoker:2001qqx, DHoker:2001foj, DHoker:2001jaf, DHoker:2005dys, DHoker:2005vch, DHoker:2007csw, DHoker:2002hof}, with the exception of certain supersymmetry-protected terms,\footnote{Notably, these include the 3-loop contribution to $D^6 R^4$ effective coupling \cite{Green:2005ba} computed in the pure spinor formalism \cite{Berkovits:2000wm, Berkovits:2006vi, Gomez:2013sla}, and higher loop contributions to F-terms in compactifications using topological strings \cite{Antoniadis:1993ze, Bershadsky:1993cx, Huang:2006hq}.} along with pieces of D-instanton corrections up to next-to-leading order \cite{Green:1997tv, Balthazar:2019rnh, Sen:2020cef, Sen:2020ruy, Sen:2021qdk, Sen:2021tpp, Sen:2021jbr, Alexandrov:2021shf, Alexandrov:2021dyl, upcoming:abcry}. Moreover, the effective gravitational coupling is expected to become strong at the Planck scale, obscuring the interpolation from perturbation theory at low energies to effective descriptions of very high-energy excitations as spacetime geometry with horizons.  In M-theory, due to the lack of a good worldvolume formulation analogous to the string worldsheet, not much is known about the scattering amplitudes of the gravitons and their superpartners beyond a few supersymmetry-protected terms in the low-energy expansion \cite{Green:1997as,Russo:1997mk,Green:1999pu,Green:2005ba}.

Despite the lack of a systematic framework, the non-perturbative S-matrices of string theory and M-theory in asymptotically flat spacetimes are nevertheless believed to exist.\footnote{Barring infrared issues in low numbers of asymptotic dimensions \cite{Strominger:2017zoo}.} A relatively novel promising direction to access them uses conformal bootstrap techniques, which are applied as follows.  First, after replacing flat space with weakly-curved AdS space, it was shown in \cite{Polchinski:1999ry,Susskind:1998vk,Giddings:1999jq,Penedones:2010ue,Fitzpatrick:2010zm} that it is possible define AdS boundary observables that are analogs of the flat space scattering amplitudes, and which approach the flat space scattering amplitudes when the curvature radius of AdS is taken to infinity.  In turn, these boundary observables can be rephrased through gauge/gravity duality \cite{Maldacena:1997re,Gubser:1998bc,Witten:1998qj}  as correlation functions in the dual, strongly coupled, large $N$ superconformal field theories, which are amenable to the conformal bootstrap program.  

To learn about string theory or M-theory using holographic CFTs, one can get the most mileage from studying maximally supersymmetric or nearly-maximally supersymmetric examples, such as: the 4d $SU(N)$ ${\cal N} = 4$ super-Yang-Mills (SYM) theory, which is dual to type IIB string theory on $AdS_5 \times S^5$;  the 6d $(2, 0)$ theory, which is dual to M-theory on $AdS_7 \times S^4$;  and the 3d $U(N)_k \times U(N)_{-k}$ Aharony-Bergman-Jafferis-Maldacena (ABJM) theory \cite{Aharony:2008ug} or the more general $U(N)_k \times U(M)_{-k}$, $N \neq M$, Aharony-Bergman-Jafferis (ABJ) theory \cite{Aharony:2008gk} --- with ${\cal N} = 8$ superconformal symmetry when $k=1, 2$ and $N=M$, and ${\cal N} = 6$ superconformal symmetry otherwise --- which are dual to M-theory on $AdS_4 \times S^7 / \mathbb{Z}_k$.  
One reason why these holographic SCFTs are the most promising bootstrap candidates is that the explicit Lagrangian descriptions in the 3d and 4d examples and the large amount of supersymmetry in all cases provide  exact results that can be derived from supersymmetric localization (see \cite{Pestun:2016zxk} for a collection of reviews and references) or from the 2d chiral algebra sectors \cite{Beem:2013sza,Beem:2014kka}, and these results can be combined with the conformal bootstrap studies.  Another reason is that the stress tensor multiplets of these maximally supersymmetric theories contain scalar operators, whose four-point functions capture the full information contained in the four-point functions of other operators in the same multiplet.  Lastly, the conformal bootstrap program constrains (S)CFTs based on symmetry and consistency conditions, so one would expect that the most constrained theories would be the most (super)symmetric ones.

 To determine the flat space S-matrix in semiclassical string/M-theory, one should extract certain coefficients in the $1/N$ expansion of correlation functions in the SCFTs mentioned above.  One approach is to first bootstrap the full four-point function of the stress tensor multiplet superconformal primary at finite $N$, and then consider the $1/N$ expansion of this result.  Another approach is to study the coefficients of the $1/N$ expansion order by order using bootstrap techniques.  As we now review, there has been progress in using the first approach numerically and the second approach analytically.

\subsubsection*{What numerical bootstrap can do for holographic theories}

\begin{figure}[h!]
\begin{center}
        \includegraphics[width=0.75\textwidth]{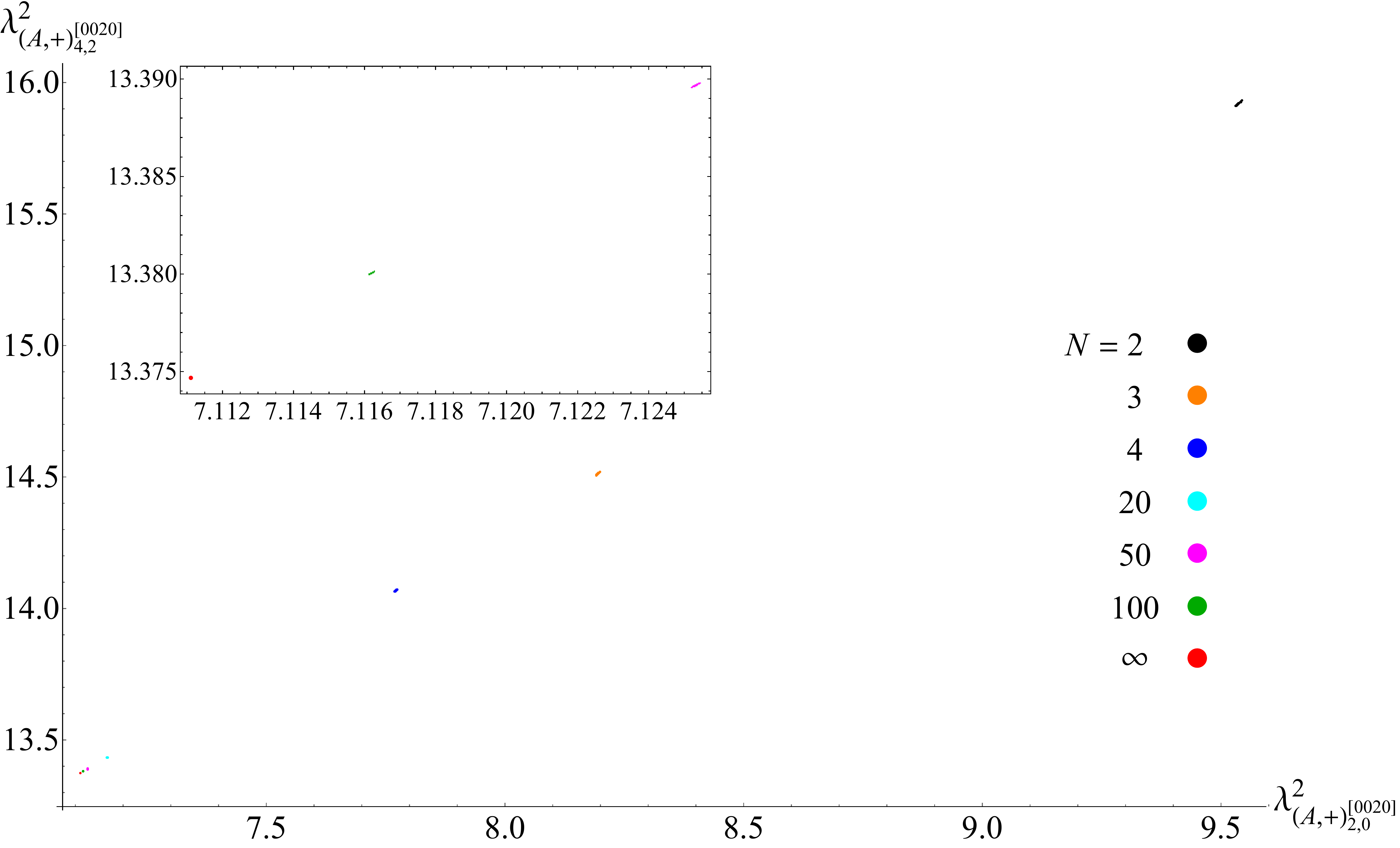}
\caption{Allowed region for OPE coefficients of operators appearing in the OPE of two stress tensor multiplet superconformal primaries for various values of $N$ for the $U(N)_1 \times U(N)_{-1}$ ABJM theory.  Figure from \cite{Agmon:2019imm}.}
\label{Archipelago}
\end{center}
\end{figure}  

The numerical bootstrap effort uses the tools being developed in the broader conformal bootstrap program for CFTs in general spacetime dimensions (for reviews and references, see \cite{Rychkov:2016iqz,Simmons-Duffin:2016gjk,Poland:2018epd,Chester:2019wfx,Qualls:2015qjb}), in particular testing for the unitarity and crossing symmetry of four-point functions using semi-definite programming, numerically implemented, for instance, with the SDPB semi-definite program solver \cite{Landry:2019qug}.    Numerical studies in the holographic theories mentioned above have been performed for 4d ${\cal N}=4$ SYM in \cite{Beem:2013qxa,Beem:2016wfs,Bissi:2020jve,Chester:2021aun},  for 3d ABJ(M) theory in  \cite{Chester:2014fya, Chester:2014mea, Agmon:2017xes, Agmon:2019imm,Binder:2020ckj}, and for the 6d $(2,0)$ theory in \cite{Beem:2015aoa}, and they include information obtained using supersymmetric localization \cite{Pestun:2016zxk,Binder:2019jwn,Chester:2020dja,Chester:2019jas,Chester:2020vyz,Dorigoni:2021bvj,Dorigoni:2021guq}.   Of the results obtained thus far, we would like to highlight the precise islands in OPE coefficient space in 3d ${\cal N} = 8$ ABJM theory (see Figure~\ref{Archipelago}), which represent a first step in determining the low-lying CFT data of ABJM theory numerically.

In future numerical studies, it will be important to make use of all information about protected correlation functions that can be obtained using supersymmetric localization.  Some of this information comes in terms of integrated correlators, which have been used in numerical bootstrap studies in more than two spacetime dimensions only recently \cite{Chester:2021aun} (see also \cite{Lin:2015wcg} for a 2d example).  It will also be important to restrict the space of theories being studied to those with a single stress tensor operator.  This can be done by analyzing the consistency of mixed systems of correlators involving, for instance, the operators studied in \cite{Agmon:2019imm,Bissi:2020jve,Bissi:2021hjk}.   With these improvements, the goal is to obtain small allowed islands in all holographic 3d and 4d SCFT examples mentioned above, to accurately extract the low-lying CFT data as function of $N$, and then to expand the result in $1/N$.

\subsubsection*{Analytic bootstrap and the low-energy (but not necessarily perturbative) expansion}

As mentioned above, another approach to large-$N$ holographic SCFTs is to bootstrap the contribution to SCFT correlators coming from bulk interactions with a small number of derivatives using consistency conditions instead of performing direct computations of Witten diagrams.  This approach is perturbative in the derivative expansion, but it is not necessarily perturbative in the string coupling constant.  In 4d ${\cal N} = 4$ SYM, for instance, this approach can be used both in the 't Hooft double scaling limit, where $N$ is taken to infinity at fixed 't Hooft coupling $\lambda = g_\text{YM}^2 N$, at large $\lambda$, as well as in the more nontrivial limit where $N$ is taken to infinity while the complexified gauge coupling $\tau$ is held fixed  \cite{Banks:1998nr,Azeyanagi:2013fla,Chester:2019jas,Chester:2020vyz}.
  
Indeed,
as initiated in \cite{Rastelli:2017udc,Rastelli:2016nze} and further developed in \cite{Alday:2017xua,Aprile:2017bgs,Alday:2017vkk,Chester:2018dga,Chester:2018aca,Binder:2018yvd,Alday:2018kkw,Alday:2018pdi,Chester:2019pvm,Binder:2019jwn,Chester:2019jas,Binder:2019mpb,Aprile:2019rep,Alday:2019nin,Alday:2020tgi,Chester:2020vyz,Alday:2021vfb,Alday:2021ymb} (see also earlier works \cite{Goncalves:2014ffa,Alday:2014tsa}), it can be shown that crossing symmetry, supersymmetric Ward identities, the Lorentzian inversion formula, as well as analytic properties of Witten diagrams in Mellin space determine the four-point functions of $1/2$-BPS external operators, up to a few undetermined constants whose number increases with the number of derivatives in the interaction vertices.  At the first few orders in the derivative expansion, these constants can be fully determined using exact results from supersymmetric localization or protected sectors. By taking the large radius limit of AdS, one can connect these results for holographic correlators  to scattering amplitudes in flat space \cite{Polchinski:1999ry,Susskind:1998vk,Giddings:1999jq,Penedones:2010ue,Fitzpatrick:2010zm}, expanded at small momentum.  This method has reproduced known results \cite{Green:1997tv,Green:1997as,Green:1998by,Green:1999pu} for the contribution of protected higher-derivative contact vertices to the string/M-theory S-matrix, for instance the supersymmetrized $R^4$ interaction \cite{Green:1997tv,Green:1997as,Green:1998by, Paulos:2008tn} whose coefficient, as a function of $\tau$, takes the form of a non-holomorphic Eisenstein series.
 
 Looking forward, one should not expect this analytic bootstrap method with input from supersymmetry-protected observables to give information about unprotected flat space interactions.  To go beyond the protected terms, one would likely have to combine the analytic bootstrap approach with other inputs coming, for instance, from the numerical bootstrap approach mentioned above.

\subsubsection*{Targets for the near future}

A concrete target for the near future would be to determine, or at least bound, the coefficient with which the first unprotected interaction term, namely the $16$-derivative interaction of the schematic form $D^8 R^4$ (together with its supersymmetric completion), contributes to the graviton S-matrix in string theory and M-theory.  In string theory, this term contributes with an unknown coefficient that is a function of $\tau$ in type IIB string theory and a function of $g_s$ in type IIA\@.  In M-theory, the coefficient of $D^8 R^4$ is just a number.\footnote{In \cite{Russo:1997mk}, it was conjectured that this term is absent in M-theory.} A more ambitious goal would be to determine the AdS analog of the Virasoro-Shapiro amplitude, i.e.~the stress tensor multiplet four-point function, as an expansion in $g_\text{YM}$ but to all orders in $1/N$ --- this would be a landmark achievement, as it would constitute a partial but substantial solution of planar $\cN=4$ SYM. (See \cite{Aprile:2020mus,Abl:2020dbx} for work in this direction.)

Another goal is to determine higher-derivative contributions to the four-point functions of the operators dual to the full Kaluza-Klein tower of modes on $AdS_5 \times S^5$, $AdS_4 \times S^7$, or $AdS_7 \times S^4$.  In tree-level supergravity, this was determined in \cite{Rastelli:2016nze,Rastelli:2017udc,Caron-Huot:2018kta,Alday:2020dtb,Alday:2020lbp} and progress has been made beyond this order for instance in \cite{Aprile:2017qoy,Alday:2018pdi,Binder:2019jwn,Drummond:2019odu,Drummond:2020uni}.  Since at a given order in the derivative expansion, the correlators corresponding to the whole Kaluza-Klein tower have their origins in the same higher derivative terms in 10 or 11 dimensions, one should search for a compact way of encoding this information. The hidden conformal symmetry of $AdS_5 \times S^5$ supergravity \cc{Caron-Huot:2018kta} and other conformally flat AdS supergravity backgrounds \cc{Rastelli:2019gtj,Abl:2021mxo} is one version of such a formulation, the deeper meaning of which is important to understand. 

In a different direction, the superconformal bootstrap may also be used to target various questions about the formal structure of string amplitudes. Number-theoretic properties of string amplitudes are a topic of active investigation in the string theory community: in particular, understanding the transcendentality properties of string amplitudes at low genus \cc{DHoker:2019blr, DHoker:2021ous}, and constraining the types of $SL(2,\mathbb{Z})$-invariant functions that can appear in both integrated \cc{Green:2008bf, Green:2014yxa, Wang:2015jna, Wang:2015aua} and unintegrated \cc{DHoker:2015gmr, DHoker:2015wxz, Brown:2017qwo, Gerken:2020yii} worldsheet amplitudes. The conformal bootstrap may be able to lend novel perspectives, via the properties of correlation functions in (say) $\mathcal{N}=4$ SYM and their flat space limits.

\subsubsection*{Defects and branes}

The dynamics of branes and open strings, beyond the low-energy effective action, can also be accessed via bootstrapping holographic theories. One approach is through SCFTs with flavor symmetry realized through flavor branes in the bulk. In this context, correlators of the conserved current multiplets are related to open string scattering amplitudes of the gluons living on the branes \cite{Alday:2021odx,Alday:2021ajh,Alday:2022lkk}. Another approach is through defect CFTs that are dual to wrapped branes, where correlators of local operators in the presence of the defect are related to closed string amplitudes in the presence of the brane. A first step in tackling this problem would be to understand whether an analog of the Mellin representation holds for correlators in the presence of defects; see \cc{Alday:2020eua} for an analogous representation for holographic thermal correlators. 

The characterization of membranes in M-theory beyond the low-energy limit has been a longstanding open question. Alternatively to the holographic approach, it is likely that the strategy for bootstrapping the S-matrix of Nambu-Goldstone modes of the flux tube \cite{EliasMiro:2019kyf} can be extended to $2+1$ dimensions with nonlinearly realized target space super-Poincar\'e symmetry, and used to constrain the dynamics of M2-branes at higher orders in the momentum expansion.

\section{AdS vacua and string universality}
\label{LANDSCAPE}

Moving away from specific top-down instances of AdS/CFT, bootstrap methods can be used more broadly to study the landscape of AdS string vacua in string theory and M-theory.  One of the main questions in pursuing such an approach is to understand the necessary and sufficient conditions for a CFT to be ``holographic,'' a colloquial shorthand to mean that the AdS dual is classical, weakly curved, and contains Einstein gravity (or a supersymmetrization thereof) as a low-energy limit. This notion originated with the conjecture of Heemskerk, Penedones, Polchinski, and Sully \cc{Heemskerk:2009pn} that there are, essentially, only two such conditions: a large central charge, and a parametrically large spectral gap to the lightest single-trace operator of spin greater than two. A significant step was taken toward its proof in \cc{Camanho:2014apa}, where graviton  three-point vertices (stress tensor three-point functions in the dual CFT language) were shown to be parametrically suppressed by the higher spin scale, $\Delta_{\rm gap}$, in powers according to dimensional analysis. This was subsequently proven and generalized to other OPE structures from conformal bootstrap points of view \cc{Afkhami-Jeddi:2016ntf, Caron-Huot:2017vep, Kulaxizi:2017ixa,  Costa:2017twz, Cordova:2017zej, Meltzer:2017rtf, Meltzer:2018tnm, Afkhami-Jeddi:2018apj, Belin:2019mnx, Kologlu:2019bco}. A key theme in those works was to study Lorentzian scattering experiments. That philosophy has recently been synthesized with a ``bottom-up'' program to constrain effective theories of gravity from analyticity and unitarity of the graviton S-matrix, leading to precise numerical bounds on coefficients in the derivative expansion of four-point amplitudes around general relativistic scattering in AdS \cc{Caron-Huot:2021rmr, Caron-Huot:2021enk}.  

A proof of the conjecture of \cc{Heemskerk:2009pn} is still lacking in complete generality, and there are some interesting open questions. How strong of an assumption is large $c$ factorization? Are there bounds on spectral data in the scalar sector, so far inaccessible to Lorentzian techniques which rely on spin? What would it take to show that arbitrary $n$-point graviton scattering matches that of general relativity at large $\Delta_{\rm gap}$?

The conditions mentioned above for a CFT to be holographic are widely believed to hold, partly because they yield predictions consistent with the Regge structure of string excitations and with the empirical properties of known top-down instances of AdS/CFT\@. This begs the question of {\it string universality}, namely, whether every consistent theory of quantum gravity is a string theory. This is a central question for modern-day research on quantum gravity. A related question from the conformal bootstrap point of view is to what extent, and under what conditions, CFTs exhibit structures characteristic of string/M-theory.\foot{This question can be further refined for the prototypical holographic CFT, which is defined by a sequence of CFTs with increasing and unbounded central charge labeled by some parameter, such as the rank of a gauge group. In this case, one can sensibly ask whether the ground state of every such CFT is dual to an AdS vacuum of string/M-theory, which can be constructed geometrically in the parametric limit. It is hard to state precisely how this notion would extend to the more general case where the CFT in question is not, in any obvious way, a member of such a sequence, though the spirit of the idea may still be meaningful.} See, for instance, \cc{Brower:2006ea, Costa:2012cb, Belin:2014fna, Haehl:2014yla, Caron-Huot:2016icg, Caron-Huot:2017vep, Sever:2017ylk, Arkani-Hamed:2020blm, Guerrieri:2021ivu, Bern:2021ppb} for varied progress in this direction. For holographic CFTs, the specific role of the higher-spin scale $\Delta_{\rm gap}$ in suppressing bulk higher-derivative interactions \cc{Camanho:2014apa} implies the necessity of higher-spin excitations. A more precise version of this connection with string theory applies to all CFTs, which have been shown to organize their spectra into families of operators that are analytic in spin, furnishing CFT analogs of Regge trajectories \cc{Caron-Huot:2017vep}. These stringy properties complement direct approaches to bootstrapping quantum gravity amplitudes \cc{Guerrieri:2021ivu, Bern:2021ppb}, in which the four-point graviton amplitudes of string theory are found to lie at very special points, close to the boundaries, in the general parameter space allowed by unitarity and crossing symmetry.

Future work is expected to address more ambitious questions. One of these is to definitively rule in/out scale-separated AdS vacua in classical string theory. This is a long-standing question in the swampland program, with differing opinions on the validity of existing string/M-theory solutions in the mold of \cc{Kachru:2003aw, Balasubramanian:2005zx, DeWolfe:2005uu}. A scale-separated AdS solution exhibits a parametric hierarchy $L_{\rm AdS} \gg L_{\mathcal{M}}$, where $\mathcal{M}$ comprises the internal dimensions of the string/M-theory compactification; the CFT dual of a putative scale-separated solution would contain a very sparse spectrum of single-trace operators with $\Delta \sim \mathcal{O}(1)$, dual only to low-lying bulk ``moduli'' rather than Kaluza-Klein modes on $\mathcal{M}$. (For preliminary work approaching this from the CFT perspective, see e.g. \cite{Aharony:2008wz,deAlwis:2014wia,Alday:2019qrf,Conlon:2021cjk,Apers:2022zjx}.) The scale separation question may be viewed as a refinement of the conjecture of \cc{Heemskerk:2009pn} that specifies the number of large bulk dimensions in which the theory exhibits sub-AdS locality. It is natural to study a version of this question in the presence of supersymmetry: if a holographic CFT possesses a continuous R-symmetry, it is not established whether this symmetry must, in fact, be geometrized at the AdS scale, notwithstanding standard AdS/CFT lore. This question is unanswered even for 4d $\cN=4$ superconformal theories: can a ``pure $AdS_5$'' compactification of string theory --- with a 5d maximal supergravity multiplet, but no large extra dimensions --- and its exotic, non-SYM $\cN=4$ SCFT dual be ruled out? (See \cc{Papadodimas:2009eu,Bonetti:2018fqz} for related comments and some possible hints.) The conformal bootstrap can rephrase \cite{Alday:2019qrf} these questions in terms of bounds on the parametric density of single-trace light operators at large $\Delta_{\rm gap}$, which could plausibly be derived by clever application of UV-IR relations, or by stronger use of the structure of Regge trajectories in large $N$ CFTs. An extreme example of this type would be a theory of ``pure gravity'' in $AdS_3$; the precise definition of such a theory varies according to taste, but no satisfactory example has been found (or ruled out), despite ongoing efforts \cc{Witten:2007kt, Maloney:2007ud, Hellerman:2009bu, Keller:2014xba, Collier:2016cls, Afkhami-Jeddi:2019zci, Hartman:2019pcd, Benjamin:2020mfz, Maxfield:2020ale, Afkhami-Jeddi:2020ezh, Maloney:2020nni}. 

Venturing one step further from the lamppost, conformal bootstrap tools will ideally be developed to allow a scan of sporadic AdS vacua of string theory. These sporadic vacua would be dual to isolated points in the space of CFTs, theories with a large but finite number of degrees of freedom, that are {\it not} members of a parametric sequence of CFTs of increasing central charge. Only then can the conformal bootstrap properly begin to address the (AdS) landscape question in string theory. Such a development could plausibly follow from new approaches to motion in the space of solutions to crossing symmetry \cite{Mazac:2019shk,Reehorst:2021ykw,Afkhami-Jeddi:2021iuw}, or from continuing the parallel work directly in AdS of first principles constraints on semiclassical gravitational scattering.

\section{The worldsheet theory: CFT and beyond}\label{WORLDSHEET}

The existence of a two-dimensional theory of gravity on the string worldsheet, commonly referred to as the ``worldsheet CFT" by a slight abuse of terminology, has been nothing short of a miracle. Here we would like to dwell and reflect on the power and limitations of the worldsheet formalism, which are not yet fully understood.

The Neveu-Schwarz-Ramond formalism \cite{Friedan:1985ge, DHoker:1988pdl, Sen:2014pia, Sen:2015hia, deLacroix:2017lif} gives a complete formulation of superstring perturbation theory\footnote{Notably, the picture changing operator formulation is completed through vertical integration \cite{Sen:2014pia, Sen:2015hia}, which allowed for a consistent formulation of perturbative NSR superstring field theory \cite{deLacroix:2017lif}. } in vacua of purely Neveu-Schwarz (NS) type. One generally expects deformations of background NSNS fields to correspond to deformations of the worldsheet ``matter" CFT, although this is only precisely formulated in time-independent backgrounds with well-defined asymptotic states. Turning on Ramond (R) sector background field strengths necessarily couples the matter and ghost systems on the worldsheet, and furthermore deforms the worldsheet CFT in a non-local manner \cite{Berenstein:1999jq, Berenstein:1999ip, Cho:2018nfn}. The pure spinor formalism \cite{Berkovits:2000wm, Berkovits:2004xu}, based on an entirely different matter and ghost structure, treats NS and R sectors on equal footing; a complete formulation of string perturbation theory in this framework remains an outstanding challenge.\footnote{In Minkowskian spacetime, the full space of states and consistency of the pure spinor worldsheet CFT remains to be clarified \cite{Berkovits:2006vi}. There appears to be no such obstacle in $AdS_5\times S^5$, where the worldsheet CFT is expected to be well-defined at least perturbatively in $\A'$ (see \cite{Bedoya:2010av} for effort in this direction). Let us remark that the Green-Schwarz formalism \cite{Green:1987sp, Green:1987mn}, in contrast, should be viewed as a construction of the string effective action that nonlinearly realizes spacetime supersymmetries, and does not a priori offer an unambiguous quantization scheme beyond leading orders in the derivative expansion unless supplemented with additional symmetries such as integrability \cite{Bena:2003wd}.}

An important class of weakly coupled strings are flux tubes in confining large $N$ gauge theories. While much has been learned about the long distance dynamics of confining strings \cite{Polchinski:1991ax, Dubovsky:2012sh, Aharony:2013ipa, Hellerman:2014cba, Dubovsky:2015zey, EliasMiro:2019kyf}, the nature of a UV-complete worldsheet theory that allows for a string theoretic description of glueball states \cite{Dubovsky:2016cog, Dubovsky:2021cor} remains unclear. The known superstring backgrounds that admit confining fundamental strings, typically acquiring spacetime geometry of the form
\be
ds^2 = f(y) dx^\mu dx_\mu + g_{ij}(y) dy^i dy^j \,,
\ee
where the warp factor $f(y)$ reaches a positive minimal value at a finite location on the base manifold (parameterized by $y^i$), necessarily involve RR flux \cite{Polchinski:2000uf, Klebanov:2000hb, Maldacena:2000yy}. It begs the question of whether a worldsheet CFT, in the sense of a 2d quantum field theory obeying locality axioms including OPE and modular invariance, really exists in this setting. The pure spinor formalism strongly hints at an affirmative answer, provided that one suitably relaxes standard assumptions on 2d unitary CFTs. To find the appropriate bootstrap axioms for the general worldsheet ``conformal theory", bridging the gap between recent efforts in bootstrapping unknown 2d CFTs \cite{Collier:2016cls, Cho:2017fzo} and effective string theories \cite{EliasMiro:2019kyf}, is likely the key to making progress on this front.

The gauge/gravity duality suggests an alternative, constructive path toward a worldsheet theory, in which the string worldsheet emerges from large $N$ gauge theories through Feynman diagrams and the spin chain structure of single-trace operators.
Recent work on concrete AdS/CFT dual pairs in the weak coupling limit of the spacetime CFT$_d$ provides a tractable testing ground \cc{Eberhardt:2018ouy, Eberhardt:2019ywk, Gaberdiel:2021jrv, Gaberdiel:2021qbb}. 
In particular, the tensionless string worldsheet theories dual to both the 2d symmetric product orbifold CFT and free ${\cal N}=4$ SYM appear to have free field descriptions \cc{Eberhardt:2018ouy, Dei:2020zui, Gaberdiel:2021jrv}. In the AdS${}_3$/CFT${}_2$ case, the string worldsheet is seen to emerge \cc{Gaberdiel:2020ycd} from the ``Feynman diagrams" of the CFT$_2$ \cc{Pakman:2009zz} very much as per the general program of \cc{Gopakumar:2003ns, Gopakumar:2004qb, Gopakumar:2005fx}, particularly the construction based on Strebel lengths in \cc{Gopakumar:2005fx, Razamat:2008zr}. This utilizes the natural appearance of the moduli space of Riemann surfaces from the 't Hooft large $N$ Feynman diagrams \cc{Gopakumar:2004qb, Gopakumar:2005fx, Razamat:2008zr}. 

The tensionless string limit for ${\rm AdS}_3$ reveals a further striking relation between the spacetime and the worldsheet, in which the latter plays the role of a covering space for the former \cc{Eberhardt:2019ywk}, concretely realizing the structure of meromorphic branched coverings {\it \`a la} Lunin-Mathur \cc{Lunin:2000yv, Pakman:2009zz} within the spacetime CFT itself. Recently, an interesting higher-dimensional analog is found that involves holomorphic covering maps into (ambi-)twistor space  \cc{Bhat:2021dez}. All of this hints at the possibility of making direct contact between conformal data of the worldsheet and spacetime CFTs, namely the operator spectrum and structure constants. Let us sketch a way to frame this connection concretely and in broader generality. 

Suppose a large $N$ gauge theory has a weakly coupled dual bulk string theory that admits a worldsheet CFT description, with the spectrum of single-trace operators mapped to single string states. The thermal partition function of the gauge theory, at temperature $T=1/\beta$ below the Hawking-Page transition, is expected to admit a well-defined infinite $N$ limit (see \cite{Sundborg:1999ue, Aharony:2003sx} for a discussion, especially in the context of weakly coupled SYM)
\ie\label{zone}
Z(\beta) = \lim_{N\to \infty} {\rm Tr}_{\rm gauge} \left[ e^{-\beta H} \right]
\fe
that counts multi-string states in the bulk as a free Fock space of single string states.  The partition function $Z(\beta)$ should be equivalently computed by the path integral over genus one string worldsheets that wind around the thermal circle, and it should admit an expression in terms of worldsheet CFT data, at least in purely NSNS backgrounds with light-like isometries \cite{Maldacena:2000kv, Eberhardt:2018ouy}, of the form\footnote{In particular, the single-trace contribution to the gauge theory partition function can be extracted from the trace over winding number 1 states in the worldsheet theory \cite{Maldacena:2000kv}.}
\ie\label{ztwo}
\log Z(\beta) =  \int_{\cal F} \frac{d^2\tau}{2\tau_2}{\rm Tr}_{\rm WS}\left[ e^{2\pi i (\tau L_0 -\bar\tau \bar{L}_0)} \right] \,.
\fe
Here, the trace on the RHS is taken over worldsheet CFT states with only transverse excitations, and the moduli integration is performed over the fundamental domain of $PSL(2,\mathbb{Z})$. The compatibility of (\ref{zone}) and (\ref{ztwo}) amounts to highly nontrivial constraints on the spectra of the two sides of the duality. An expression analogous to (\ref{ztwo}) should hold whenever a worldsheet CFT exists, and may serve as the starting point for bootstrapping this worldsheet theory. 

Even richer constraints remain to be disentangled from the data of the three-point functions. In Minkowskian spacetime, the unitarity and locality property of perturbative string amplitudes are closely tied to the OPE in the worldsheet CFT\@. For string theories in AdS$_{d+1}$, one might expect a similar relationship between the OPE of single-trace operators in the dual spacetime CFT$_d$ and some suitable notion of OPE in the worldsheet theory \cite{Aharony:2007rq, Ghosh:2017lti}. 
It remains a major challenge to identify the appropriate bootstrap conditions on the putative worldsheet OPE\@. The task is to recast these ``dual bootstrap" constraints into an effective form like in the conventional bootstrap.  A concrete aim would be to obtain the worldsheet description of weakly coupled ${\cal N}=4$ SYM using this approach.

\section{Longer-term targets}\label{POST}

The next generation of bootstrap techniques will require new ideas. Let us mention a few possible avenues that may be relevant for string theory. In the context of the superconformal bootstrap, one would like to acquire a better understanding of the implications of (generalized) S-duality, for $\cN=4$ SYM and other SCFTs \cc{Montonen:1977sn,Witten:1978mh,Osborn:1979tq,Argyres:2006qr,Argyres:2007cn}. Many SCFTs have exactly marginal couplings on which S-duality symmetries act; in holographic duality with string theory, these are identified with U-duality symmetries preserved by an AdS compactification. The bootstrap program has thrived in part because of the efficient progress in identifying and processing symmetries: for example, computations of conformal blocks and Ward identities, and the discovery of unobvious symmetries \cc{Beem:2013sza,Beem:2014kka, Beem:2014rza, Chester:2014mea,Beem:2016cbd,Caron-Huot:2018kta}. To extract the physical consequences of S-duality for CFT data and the AdS dual string quantities, the bootstrap should incorporate these symmetries as efficiently as possible (see \cc{Collier:2022emf} for a new approach). Similar comments apply to integrability and emergent Yangian symmetry at large $N$: for $\cN=4$ SYM \cc{Beisert:2010jr} and other maximally supersymmetric SCFTs \cc{Babichenko:2009dk,Borsato:2014hja}, the techniques and results of planar integrability should be combined with, or fed into, the conformal bootstrap at large $N$. 

\begin{figure}[t]
\begin{center}
        \includegraphics[width=0.8\textwidth]{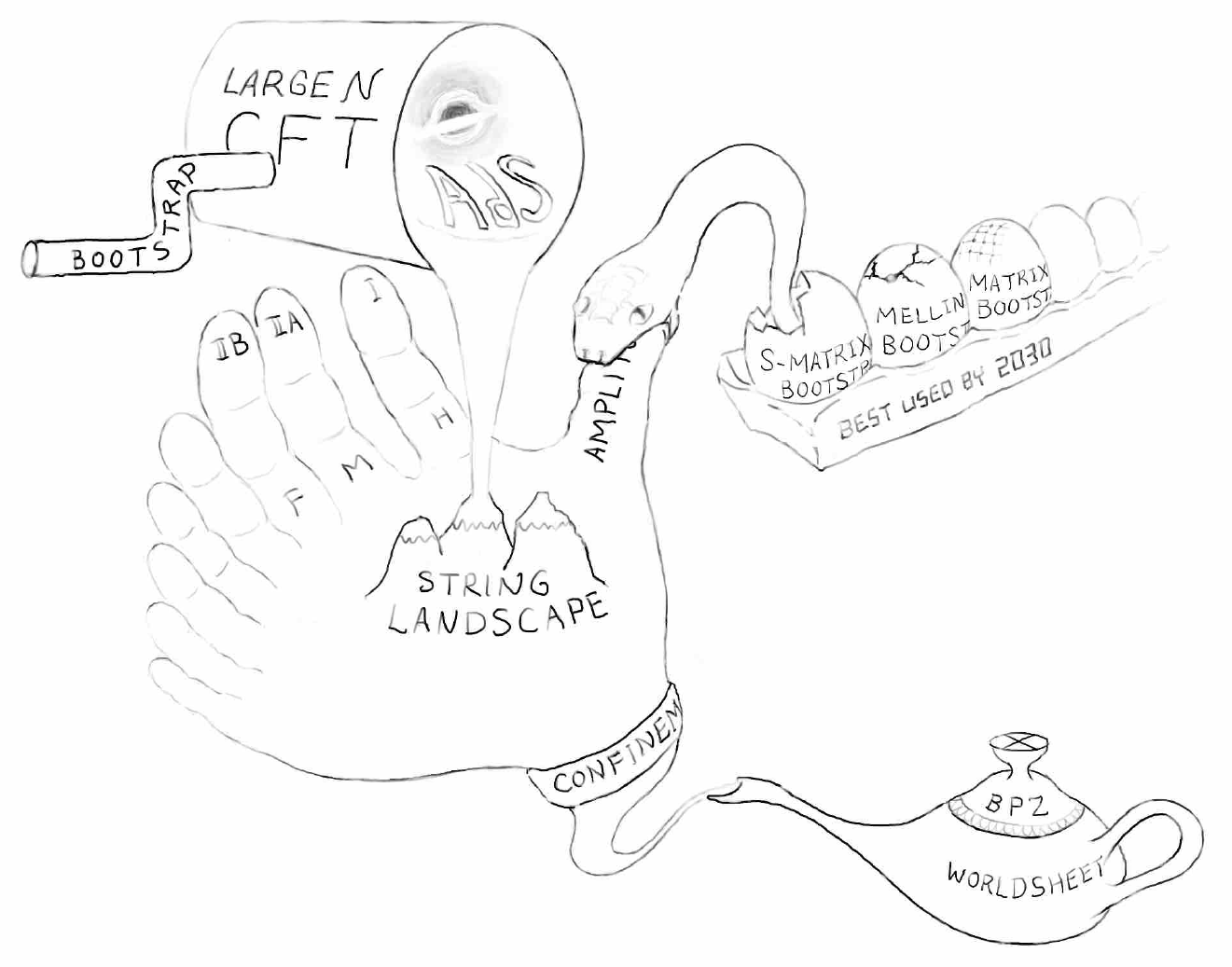}
\label{art}
\end{center}
\end{figure}  

We close this paper by describing some research directions that are further afield; speculations on the future; and a visual depiction of the past, present, and possible future of bootstrapping string theory. 

\subsubsection*{Beyond the low-energy expansion}

Current efforts in bootstrapping supersymmetric holographic CFTs largely focus on the four-point correlator of BPS multiplets. The latter, being merely an analytic function of the cross ratios $z,\bar z$ and dependent on the central charges and couplings, in principle contains a huge amount of information including graviton $2\to 2$ S-matrix element at all energies and impact parameters, capturing Planckian and black hole physics. While analytic bootstrap efforts have concentrated on the lightcone limit \cite{Fitzpatrick:2012yx, Komargodski:2012ek, Kaviraj:2015cxa, Alday:2015ewa, Alday:2016njk, Simmons-Duffin:2016wlq,Caron-Huot:2017vep} and the Regge limit \cite{Costa:2012cb, Kulaxizi:2017ixa, Li:2017lmh},\footnote{The lightcone limit amounts to taking $z\to 0$, fixing $\bar z$. The Regge limit is defined as $z\to 0$, $z/\bar z$ fixed, after performing a monodromy of $\bar z$ around 1 on the complex plane. The bulk point limit is defined by taking $z\to \bar z$ after performing the same monodromy of $\bar z$ around 1. Euclidean correlators, on the other hand, correspond to complex conjugate $z,\bar z$.} numerical bootstrap results that determine internal operator dimensions and OPE coefficients within small islands are a priori applicable in the Euclidean regime where the conformal block expansion converges absolutely. To access high-energy scattering in the flat space limit requires taking the bulk point limit, and at the same time $N\to \infty$ with finite (gauge) coupling. Whether the bootstrap can pin down correlators in the bulk point limit, possibly through high precision numerics, remains a major challenge.

Another trans-Planckian phenomenon of interest in string theory is the annihilation/decay of branes \cite{Alexander:2001ks, Kachru:2002gs, Sen:2002nu}. A possibly tractable example through holography is the annihilation of a pair of giant gravitons in $AdS_5\times S^5$ \cite{McGreevy:2000cw} into closed string radiation. Even though such amplitudes are in principle related to the scattering of closed strings off a D-brane via analytic continuation and crossing, the physical process occurs at an energy scale far beyond the realm of string perturbation theory. The answer is in principle contained in correlation functions involving determinant operators (see \cite{Jiang:2019xdz} for recent progress on their OPE coefficients) and may be amenable to the bootstrap. Vis-\`a-vis string universality, a more abstract goal is to bootstrap the necessity of branes themselves.

\subsubsection*{Thermal bootstrap, matrix bootstrap, and black holes}

So far, the conformal bootstrap has had comparatively little to say about finite temperature physics \cite{Iliesiu:2018fao,Iliesiu:2018zlz}, and even less to say about thermal phases of string theory specifically. A partial exception to this is in CFT$_2$, thanks to the power of modular invariance on the torus; however, we note that the modular bootstrap has made little contact with string theory or string universality.

Perhaps the main goal is to describe the quantum states or resonances of a black hole that account for its macroscopic entropy \cite{Bekenstein:1973ur, Hawking:1976de}. The successful microscopic explanation of the entropy for certain supersymmetric black holes \cite{Strominger:1996sh, Benini:2015eyy, Benini:2016rke, Cabo-Bizet:2018ehj, Choi:2018hmj, Gaiotto:2021xce} demonstrates convincingly that string theory and holographic dualities capture the fundamental degrees of freedom of quantum gravity. Far less is understood concerning far-from-supersymmetric black holes, with the notable exception of results from Monte Carlo simulation of strongly coupled matrix quantum mechanics \cite{Anagnostopoulos:2007fw}. Recently, a new bootstrap based on loop equations and positivity constraints \cite{Anderson:2016rcw, Lin:2020mme, Han:2020bkb, Kazakov:2021lel, KITPtalks} has been successfully applied to large $N$ matrix models. While still in its infancy, this approach opens a new possible pathway toward probing black hole states from strongly coupled matrix models. 

Some obvious but worthwhile targets for the conformal bootstrap are to derive the famous factor of $3/4$ relating the thermal free energies of planar $\mathcal{N}=4$ SYM at weak and strong 't Hooft coupling \cc{Gubser:1996de}, and likewise for the viscosity-to-entropy-density ratio $\eta/s=1/4\pi$ \cc{Kovtun:2004de}. Deriving either of these universal properties of gravity at long wavelengths would require extending bootstrap methods to new types of observables. 

It is also worth stressing that the role of string theory in resolving the black hole information paradox and related puzzles is not yet known. Recent computations (see \cc{Penington:2019kki,Almheiri:2019qdq,Bousso:2022ntt} and references therein) of the Page curve for certain evaporating black holes using the path integral of semiclassical gravity use surprisingly little of the microscopic details of the bulk theory; the cost is the absence of an explicit accounting of the microstates under time evolution. It would be very satisfying if a full solution of this paradox could be shown to require stringy effects, properly understood. Perhaps the conformal bootstrap at large $N$ can have something to say about this interesting problem or its AdS avatars.

\subsubsection*{The Polyakov-Mellin bootstrap and string theory}
An alternative to the conventional conformal block expansion of correlators and imposing crossing symmetry constraints was proposed long ago by Polyakov \cc{Polyakov:1974gs}.  Here one expands in a manifestly crossing-symmetric basis of functions whereupon the constraints now consist of requiring the contributions of certain spurious double-trace operators to vanish. This approach has been revisited \cc{Gopakumar:2016wkt, Gopakumar:2016cpb, Gopakumar:2018xqi} and it was realised that a) Polyakov's crossing symmetric basis for CFT$_d$ is nothing other than the all-channel sum of exchange Witten diagrams in AdS$_{d+1}$ (together with additional contact Witten diagrams that have recently been determined by consistency \cc{Gopakumar:2021dvg}); and b) the constraints are simplest to impose in Mellin space where they become the vanishing of residues of certain spurious single and double poles. These conditions are equivalent to those obtained from conventional dispersion relations for Mellin amplitudes, which now have a rigorous definition that allows for a careful specification of the Polyakov conditions (see \cc{Penedones:2019tng, Carmi:2020ekr, Caron-Huot:2020adz}). 

Both of the above features are very suggestive of a reorganisation of arbitrary CFT$_d$ amplitudes in terms of a string (string field?) theory description in AdS$_{d+1}$. It has been seen that this expansion is analytically very efficient at obtaining the first few orders results, in the $\epsilon$-expansion, for the dimensions and OPE coefficients of the $O(N)$ Wilson-Fisher fixed points \cc{Sen:2015doa, Gopakumar:2016wkt, Gopakumar:2016cpb, Dey:2016mcs}. In SCFTs, such an expansion provides a clearer interpretation of various quantities that can be computed using supersymmetric localization \cite{Binder:2021euo}.  So one might hope that this AdS reorganization of CFTs is not just a technical convenience, but pointing to a stringy picture, much as in the long sought-after string theory dual to the 3d Ising model \cc{Polyakov:1987hqn}. 

More generally, the previous paragraphs describe one specific recasting of what it means to apply the conformal bootstrap method to correlation functions, where crossing symmetry is traded for analyticity; can we find others, that can be algorithmically solved?

\subsubsection*{What is (admissible in) string theory?}

A large class of constructions of string vacua is based on F-theory compactification \cite{Vafa:1996xn, Denef:2008wq, Weigand:2018rez}, which takes into account S-duality monodromy branes of type IIB string theory and goes beyond both effective field theory and string perturbation theory. While geometric singularities play an essential role in the F-theory paradigm, it would be desirable to have a first principles justification or classification of the admissible singularities, as well as a framework for dynamics beyond what is governed by the moduli space of vacua and BPS objects. 

A basic example is D7-brane in type IIB string theory. In its presence, string perturbation theory based on a worldsheet theory with Dirichlet boundary condition breaks down, reflecting the fact that the D7-brane cannot exist on its own but must be modified into the stringy cosmic string solution \cite{Greene:1989ya}. Questions concerning the dynamics of the stringy cosmic string may be good targets for the S-matrix bootstrap.

\subsubsection*{Bootstrapping de Sitter quantum gravity}

Quantum gravity in de Sitter space, even the question of in what sense it exists if at all, remains a mystery \cite{Strominger:2001pn, Witten:2001kn, Spradlin:2001pw, Anninos:2012qw, Susskind:2021omt}. A natural set of observables in de Sitter space are the late-time correlation functions (``cosmological correlators") \cite{Strominger:2001pn, Maldacena:2002vr, Arkani-Hamed:2015bza,Arkani-Hamed:2018kmz,Pajer:2020wnj,Goodhew:2020hob,Sleight:2019hfp,Baumann:2019oyu,Baumann:2020dch,Sleight:2020obc,Sleight:2019mgd,Meltzer:2021zin} that may be amenable to a bootstrap approach. While these correlators are subject to (Euclidean) conformal symmetry, the analog of the locality/OPE property as well as the meaning of (bulk) unitarity are not understood. Recently steps in this direction have been taken in the context of quantum field theory in de Sitter space \cite{DiPietro:2021sjt, Hogervorst:2021uvp}, where the notion of OPE and bootstrap axioms are formulated. The outstanding challenge is to extend such axioms to quantum gravity.

\section*{Acknowledgements}

XY would like to thank Ying-Hsuan Lin for discussions. RG is supported in part by the J.C.~Bose National Fellowship of the DST-SERB, by the DAE project no.~RTI4001 and the broad framework of support for basic sciences from the people of India. EP is supported by ERC Starting Grant 853507. SSP is supported in part by the Simons Foundation Grant No.~488653, and in part by the US NSF under Grant No.~2111977. XY is supported by a Simons Investigator Award from the Simons Foundation, by the Simons Collaboration Grant on the Non-Perturbative Bootstrap, and by DOE grant DE-SC0007870.

\bibliographystyle{JHEP}
\bibliography{white_bib}
\end{document}